\documentclass[aip,jcp,reprint,floatfix,a4paper]{revtex4-1}

\usepackage{graphicx}
\usepackage[sort&compress]{natbib}
\usepackage[sumlimits]{amsmath}
\usepackage{amsfonts}
\usepackage{amssymb}
\usepackage{dcolumn}
\usepackage{subdepth}
\usepackage{color}
\usepackage{enumitem}
\usepackage{hyperref}
\usepackage{blindtext}
\usepackage{subdepth}

\begin{document}
\title{Quasi-classical approaches to vibronic spectra revisited}

\affiliation{Institute of Physics, Rostock University, Albert-Einstein-Str. 23-24, 18059 Rostock, Germany}
\author{Sven Karsten}
\author{Sergei D. Ivanov}
\email{sergei.ivanov@uni-rostock.de}
\author{Sergey I.\ Bokarev}
\author{Oliver K\"uhn}

\newcommand{\mOp}[1]{\hat{#1}}
\newcommand{\mVec}[1]{\boldsymbol{\mathrm{#1}}}
\newcommand{\mBra}[1]{\langle #1 |}
\newcommand{\mKet}[1]{| #1 \rangle}
\newcommand{\mOpInt}[1]{\mOp{#1}^{(0)}}
\newcommand{\mAve}[1]{\left \langle #1 \right \rangle}
\newcommand{\Eq}[1]{Eq.\,(\ref{#1})}
\newcommand{\Eqs}[1]{Eqs.\,({#1})}
\newcommand{\Sec}[1]{Sec.\,\ref{#1}}
\newcommand{\mMat}[1]{{\boldsymbol{\mathbb{#1}}}}
\newcommand{\mTen}[1]{{\boldsymbol{\mathcal{#1}}}}
\newcommand{\FT}[1]{S}
\newcommand{\e}{\mathrm{e}}
\newcommand{\im}{\mathrm{i}}
\newcommand{\diff}{\mathrm{d}}
\newcommand{\cc}[1]{\textcolor{red}{#1}} 
\newcommand{\Fig}[1]{Fig.\,\ref{#1}}
\newcommand{\eq}[1]{Eq.~(\ref{#1})}
\newcommand{\cm}{cm$^{-1}$}
\newcommand{\QCF}{SCF}
\newcommand{\QCFs}{SCFs}
\renewcommand{\Re}{\mathrm{Re}}
\newcommand{\Tr}{\mathrm{Tr}}
\newcommand{\tr}{\mathrm{tr}}
\renewcommand{\l}{\lambda}
\newcommand{\w}{\Omega}
\newcommand{\T}[1]{#1^{\mathrm{T}}}
\newcommand{\deq}{r^\mathrm{eq}}
\newcommand{\Mom}{\Pi}
\newcommand{\eps}{\varepsilon}

\begin{abstract}
The framework to approach quasi-classical dynamics in the electronic ground state is well established and is based on the Kubo-transformed time correlation function (TCF), being the most classical-like quantum TCF.
Here we discuss whether the choice of the Kubo-transformed TCF as a starting point for simulating \textit{vibronic} spectra is as unambiguous as it is for vibrational ones.
A generalized quantum TCF is proposed that contains many of the well-established TCFs as particular cases.
It provides a framework to develop numerical protocols for simulating vibronic spectra via quasi-classical trajectory-based methods that allow for dynamics on many potential energy surfaces and nuclear quantum effects.
The performance of the methods based on the well-known TCFs is investigated on 1D anharmonic model systems at finite temperatures.
The flexibility inherent to the formulation of the generalized TCF provides a route to construct new TCFs that may lead to better numerical protocols as is shown on the same models.
%
\end{abstract}

\date{\today}
\maketitle

\section{Introduction}
%
Understanding the dynamics of complex many-body systems is the grand challenge of theoretical chemistry and molecular physics.
Recent decade witnessed a spectacular progress in (non-linear) experimental spectroscopic techniques~\cite{Mukamel-Book,Kuehn-Book,Hamm-Zanni-Book} in various frequency ranges, owing to the appearance of ultra-short pulses and intense light sources.~\cite{Hwang_JMO_2015, Cho-CR-2008,Milne2014,Teichmann_NC_2016}
The resulting vibrational, electronic and vibronic spectra provide comprehensive information about the dynamical processes, when interpreted and understood with the help of proper theoretical tools.

From the theoretical standpoint, there exist two limiting strategies to simulate vibronic spectra, that is energy and time-domain approaches.
For the former, in the simplest case single-point electronic structure calculations are performed and broadening is included on a phenomenological level.~\cite{Grimme-Review-2004,Mukamel-Book}
Further, nuclear vibrations can be treated within the Franck-Condon model assuming shifted harmonic potentials for the initial and final electronic states.~\cite{Kuehn-Book, Grimme-Review-2004,Wachtler2012,Schroeter-PR-2015}
Still, this approach is not appropriate for cases where strong anharmonicities, bond formation or cleavage, and/or pronounced conformational changes are observed.
In the time domain arguably the best approach is to perform wavepacket quantum dynamics numerically exactly.~\cite{Heller1978,Heller-JCP-1979,Kuehn-PRP-2006,Wachtler2012,Schinke-Book, MCTDH-Book}
However, it usually requires an expensive pre-computation of many-dimensional potential energy surfaces (PES) and is limited to small systems or is based on a reduction of dimensionality.
Many attempts to bridge the gap between the two extrema, which are not possible to review in detail here,  were made, see Refs.~\onlinecite{Martinez-JPCA-2000,Quantics,Tavernelli-ACR-2015} and references therein for selected examples.
The consensus is that it is desirable to have a method that would combine the advantages of the two limiting strategies in an optimal way.
If one starts from the energy-domain approaches, a step in this direction is to sample nuclear distributions in the phase space via molecular dynamics (MD) methods.~\cite{Kuehn-Book, Marx-Book} 
It leads to a more realistic description of conformational and environmental effects,~\cite{Oncak-JCP-2010,Jena2015, Weinhardt2015} but still lacks information about correlated nuclear motion and thus, for instance, is not capable of reproducing vibronic progressions.
Correlations can be included by recasting the quantities of interest in terms of time correlation functions (TCFs).~\cite{lawrence2002,harder2005,Ivanov-PCCP-2013,Kuehn-Book,Mukamel-Book}

Recently, we have developed such an extension to the state-of-the-art sampling approach to X-ray spectroscopy, in particular to X-ray absorption and resonant inelastic X-ray scattering  spectra.~\cite{Karsten-JPCL-2017,Karsten-JCP-2017}
Further improvements of the method should attack the main approximations behind it: the Born-Oppenheimer approximation and the dynamical classical limit (DCL).~\cite{Kuehn-Book}
The former leads to a neglect of non-adiabatic effects, which are conventionally treated via surface hopping methods,~\cite{Tully-JCP-1971,Tully-FD-1998} mean-field (Ehrenfest) dynamics,~\cite{Tully-FD-1998} multiple
spawning techniques,~\cite{Martinez-JPCA-2000,Levine-AR-2007} classical and semiclassical mapping approaches,~\cite{Meyer-JCP-1979,Stock-PRL-1997,Bonella-CP-2001} exact factorization perspective~\cite{Gross-PRL-2010,Agostini-JCTC-2016} and Bohmian dynamics~\cite{Tavernelli-JCP-2013} to mention but few, see, e.g., Refs.~\onlinecite{Stock_Thoss-2005,Tully-JCP-2012,Tavernelli-ACR-2015} for review.
The consequences of the DCL approximation are twofold.
First, the nuclear dynamics is exclusively due to forces in the electronic ground state.
It leads to the complete loss of information about the excited state dynamics and can cause wrong frequencies and shapes of the vibronic progressions in certain physical situations, although the envelopes of the vibronic bands may be reproduced reasonably well.~\cite{Egorov-JCP-1998,Rabani-JCP-1998}
Furthermore, at ambient temperatures the ground state trajectory is confined in a small region near the potential minimum, i.e.\ within DCL one cannot describe excited-state dynamics such as dissociation.
Further, the nuclei are treated as point particles, sacrificing their quantum nature, in particular zero-point energy and tunneling effects.
This might lead to qualitatively wrong dynamics and even sampling, if light atoms, shallow PESs and/or isotope substitutions are involved, as have been shown on numerous examples starting from small molecules in gas phase to biomolecules.~\cite{Ivanov-NatChem-2010,Witt-PRL-2013,Olsson-JACS-2004,Gao-AR-2002}

In order to improve on the DCL, a method  that explicitly accounts for excited states' dynamics is needed.
Following Ref.~\onlinecite{Shemetulskis-JCP-1992}, one can derive a semiclassical approximation to the absorption cross section that leads to the dynamics that is performed on the arithmetic mean of the ground and excited state PESs,
hence referred to as the averaged classical limit (ACL) method.
Note that such a derivation for resonant Raman spectra leads to the known expression derived by Shi and Geva.~\cite{Shi-JCP-2005,Shi-JCP-2008}
The authors 
evaluated the quality of the ACL method on simple test systems and found it satisfactory.

For inclusion of quantum effects, Feynman path integrals (PI) provide hitherto the most elegant and robust solution for trajectory-based approaches.~\cite{Feynman-Book-1965,Schulman-Book,Marx-Book,Tuckerman-Book}
Here, the ring polymer molecular dynamics (RPMD) method~\cite{Craig-JCP-2004} enjoyed success in simulating quasi-classical dynamics, see e.g.\ Ref.~\onlinecite{Ceriotti-CR-2016} for review.
Further, two similar non-adiabatic versions of RPMD (NRPMD) had been developed,~\cite{Richardson-JCP-2013,Ananth-JCP-2013} based on the mapping approach.~\cite{Thoss-PRA-1999,Stock_Thoss-2005}
%
This method allows for all the aspects discussed above and is a suitable method of choice, given an efficient simulation protocol is provided.~\cite{Richardson-CP-2016}
The cornerstone of (N)RPMD is the Kubo-transformed TCF, which is the most classical-like quantum TCF, since it is real-valued and symmetric with respect to time reversal.
Nonetheless, when it comes to practical evaluation of \textit{vibronic} spectra, the Kubo TCF either becomes non-tractable by MD methods or has to be decomposed into the contributions that do not have the beneficial properties of the original TCF, in particular they are no more real functions of time, see \Sec{sec:practical_considerations} for details. 
This poses the central question of this work, that is whether the choice of the Kubo TCF as the starting point for simulating vibronic spectra is as unambiguous as it is in  infrared spectroscopy.~\cite{Craig-JCP-2004,Ramirez2004,Ivanov-PCCP-2013}

In anticipation of our results, we propose a generalized TCF that contains most of the well-established ones, including the Kubo TCF, as particular cases.
The practical recipe for simulating absorption spectra based on it employs the dynamics with respect to several PESs and incorporates nuclear quantum effects via the imaginary time PI formalism.
We demonstrate that the best results indeed do not necessarily come from the Kubo TCF.

The paper is structured as follows. 
In Sec.~\ref{sec:Theory}, a generalized TCF is introduced and approximations to it are performed in the framework of the imaginary time PI method.
The connection of the developed formalism to the well-established TCFs and methods are discussed in \Sec{sec:limiting_cases}.
The results for parametrized one-dimensional model systems (see Sec.~\ref{sec:Computational_Details}) are presented in Sec.~\ref{sec:Results} along with some recipes how to improve the numerical behavior.
Conclusions and outlook can be found in Sec.~\ref{sec:conclusions}

\section{Theory}
\label{sec:Theory}

\subsection{Generalized time correlation function}
\label{sec:Imag_time_shifted_tcf}
The experimentally measured quantity, the absorption cross-section is proportional to the lineshape function 
\begin{equation}
\FT{C}_0(\w)=\intop_{-\infty}^{\infty} \diff t \, \e^{-\im \w t} C_0(t)
\enspace,
\end{equation}
which is given by the Fourier transform of the dipole autocorrelation function~\cite{Kuehn-Book}
\begin{equation}
\label{eq:dipole_TCF}
C_0(t)\equiv \frac{1}{Z}\tr \left [ \e^{-\beta\mOp{H}}\mOp{d}(0)  \mOp{d}(t) \right  ] 
\enspace;
\end{equation}
the TCFs will be denoted with $C$ and their respective Fourier transforms with $S$ throughout the manuscript.
Here $\mOp{H}$ is the full molecular Hamiltonian of the system including both electrons and nuclei, $Z=\tr  [ \exp(-\beta\mOp{H}) ]$ is the respective partition function, $\beta\equiv1/k_\mathrm{B} T$ being the inverse temperature, and $\mOp{d}(t)$ is the total dipole operator time-evolved with respect to $\mOp{H}$.

It is well known that many quantum TCFs can be defined, all carrying the same information since their Fourier transforms have simple relations.~\cite{Ramirez2004}
In particular, applying a shift in imaginary time to the dipole autocorrelation function, \Eq{eq:dipole_TCF}, leads to
\begin{align}
\label{eq:shifted_tcf}
C_{\l}(t)\equiv C_0(t+\i\l\hbar)
=\frac{1}{Z}\tr \left [ \e^{-(\beta-\l)\mOp{H}}\mOp{d}(0) \e^{-\l\mOp{H}}  \mOp{d}(t) \right  ]
\enspace,
\end{align}
hence referred to as the imaginary time shifted TCF.
The aforementioned relation for the Fourier transforms of $C_\l$ and $C_0$ reads
\begin{equation}
 \FT{C}_\l(\w) =\e^{-\l\w\hbar} \FT{C}_0(\w)
 \end{equation}
as it is shown in the Supplement.
Integrating both sides of this equation over $\l$ from 0 to $\beta$ yields the relation between the absorption spectrum and the well-known Kubo-transformed TCF.~\cite{Kubo1957}
In an attempt to formulate a more general and flexible approach that may lead to more practical simulation protocols, we propose to employ a weighting function $w(\l)$ for this integration leading to
\begin{align}
\label{eq:main}
\underbrace{\frac{1}{\beta}\intop_{0}^{\beta} \diff \l \, w(\l) \FT{C}_\l(\w)}_{\bar{\FT{C}}_w(\w)}=
\underbrace{\frac{1}{\beta}\intop_{0}^{\beta} \diff \l \, w(\l) \e^{-\l\w\hbar}}_{p_w(\w)} \FT{C}_0(\w)
\enspace,
\end{align}
such that the absorption lineshape can be obtained as
\begin{align}
\label{eq:Sbar}
\FT{C}_0(\w)=p^{-1}_w(\w) \bar{\FT{C}}_w(\w)
\enspace.
\end{align}
Naturally, setting $w(\l)=1$ would yield back the Kubo-transformed TCF.

Given the flexibility provided by the arbitrary choice of the weighting function $w(\l)$, we attempt to find a reasonable approximation to $\bar{C}_w(t)$,
\begin{equation}
\label{eq:Cwt}
 \bar{C}_w(t) = \frac{1}{\beta}\intop_0^\beta \diff \l w(\l) C_\l (t)
 \enspace ,
\end{equation}
which is a time-domain version of $\bar{\FT{C}}_w(\w)$,
rather than to approximate the desired lineshape function, $\FT{C}_0(\w)$, directly.
The prefactor $p^{-1}_w(\w)$ in \Eq{eq:Sbar} compensates the performed shift in the imaginary time and is thus referred to as the shift correction factor (\QCF).
As it will become clear later, the \QCF\ acts as a ``magnifying glass'' or a ``filter'' emphasizing certain contributions stemming from particular $\l$ and suppressing the others.
It is worth noting that \Eq{eq:Sbar} serves as a common starting point for several popular approximations to vibronic spectroscopy as will be shown later.
%
\subsection{Practical considerations}
\label{sec:practical_considerations}
%
In order to formulate a reasonable approximation to $\bar{C}_w(t)$, the central object to consider is the imaginary time shifted TCF, $C_\l(t)$, defined in \Eq{eq:shifted_tcf}.
Assuming the Born-Oppenheimer approximation and evaluating the electronic part of the trace in \Eq{eq:shifted_tcf} in the adiabatic basis $\mKet{a}$ yields
\begin{align}
\label{eq:shifted_tcf_BO}
&C_{\l}(t)=\frac{1}{Z}\sum_{a,b}\Tr \left [ \e^{-(\beta-\l)\mOp{H}_a}\mOp{D}^{a}_b \e^{-\l\mOp{H}_b} \e^{\i\mOp{H}_b t / \hbar} \mOp{D}^{b}_a  \e^{-\i\mOp{H}_a t / \hbar} \right  ] \enspace,
\end{align}
where the capital $\Tr [\bullet]$ stands for a trace in the nuclear Hilbert space only, $\mOp{H}_a$ corresponds to the nuclear Hamiltonian with the PES of the $a$-th electronic eigenstate and  $\mOp{D}^{a}_b=\mBra{a} \mOp{d} \mKet{b}$ is the transition dipole moment.
For the sake of brevity a single transition from an initial electronic state $g$ to a final state $f$ of a two-level system is considered in the following.
The generalization to the case of many states is straightforward.

\begin{figure}[tb]
\includegraphics[width=0.9\columnwidth]{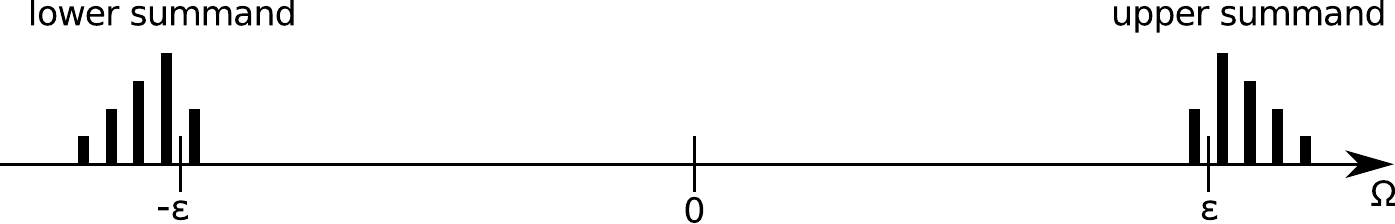}
\caption{\label{fig:spec_sketch}
Schematic picture of a spectrum resulting from the two summands in \Eq{eq:ordered_TCF_evaluated}.
Note that the \QCF, which would lead to different relative intensities at $\pm \eps$, is not applied.
}
\end{figure}

Spelling out the electronic trace in \Eq{eq:shifted_tcf_BO} leads to
\begin{align}
\label{eq:ordered_TCF}
\nonumber
C_{\l}(t) = &
\frac{1}{Z}\Tr \left [ \e^{-(\beta-\l)\mOp{H}_g}\mOp{D}^{g}_f \e^{-\l\mOp{H}_f} \e^{\i\mOp{H}_f t / \hbar} \mOp{D}^{f}_g  \e^{-\i\mOp{H}_g t / \hbar} + \right. \\
&\left. \e^{-(\beta-\l)\mOp{H}_f}\mOp{D}^{f}_g \e^{-\l\mOp{H}_g} \e^{\i\mOp{H}_g t / \hbar} \mOp{D}^{g}_f  \e^{-\i\mOp{H}_f t / \hbar} \right  ]
\enskip .
\end{align}
It is possible to shift the final state Hamiltonian as $\mOp{H}_f\equiv\mOp{\tilde{H}}_f +\eps_{fg}$ such that the eigenvalues of $\mOp{H}_g$ and shifted $\mOp{\tilde{H}}_f$ are in the same energy range and thus the respective frequencies are on the same (nuclear) timescale. 
Such a shift is needed for practical purposes, since the TCF would otherwise oscillate on the electronic timescales, which are unresolvable by means of nuclear MD methods.~\cite{Karsten-JPCL-2017,Karsten-JCP-2017}
To reiterate, the shift is equal to zero by construction when purely vibrational transitions are concerned and this issue does not occur.

Applying the shift to \Eq{eq:ordered_TCF} and evaluating the trace in the nuclear eigenstates $\mOp{H}_g\mKet{G}=E_G\mKet{G}$ and $\mOp{\tilde{H}}_f\mKet{F}=\tilde{E}_F \mKet{F}$ yield
%
%
\begin{widetext}
\begin{align}
\label{eq:ordered_TCF_evaluated}
\nonumber
C_{\l}(t) =& \frac{1}{Z} \sum_{G,F} \e^{-(\beta-\l)E_G} \e^{-\l (\tilde{E}_F+\eps_{fg})}| \mBra{G} \mOp{D}^{g}_f \mKet{F}|^2    \e^{\im (\eps_{fg}+\tilde{E}_F-E_G) t / \hbar}   \\
+& \frac{1}{Z} \sum_{G,F}  \e^{-(\beta-\l)(\tilde{E}_F+\eps_{fg})} \e^{-\l E_G} | \mBra{G} \mOp{D}^{g}_f \mKet{F}  |^2 \e^{-\im (\eps_{fg}+\tilde{E}_F-E_G ) t / \hbar}
\enskip ;
\end{align}
\end{widetext}
note the different sign in front of $\eps_{fg}$ in the phase factors.
The transitions due to the first and second summands group near the shift frequencies $\pm \eps_{fg}$.
If $\eps_{fg} \gg \tilde{E}_F-E_G$, it becomes apparent that both summands yield completely isolated spectral features around $\w=\pm \eps_{fg}$, see \Fig{fig:spec_sketch}.
This makes MD methods unsuitable for the present purpose, as the spectrum covers frequency ranges unaccessible to them.
Consequently both summands have to be considered individually in practice.
However, the TCFs described by the individual terms do not have the properties of the common TCFs described by \Eq{eq:ordered_TCF}.
For instance, the respective terms for Kubo TCF would become complex when treated this way,
as can be easily seen by choosing $w(\l)=1$.
%
Therefore the terms themselves do not constitute an obvious choice for an approximation by (quasi-)classical methods.
In a nutshell, for a practical application one can not approximate the TCF given by \Eq{eq:ordered_TCF} directly, as it requires treating the dynamics on the electronic timescales, whereas approximating the summands therein individually does not favor the Kubo-transformed TCF, because the summands are not real functions of time.
In the following only the first term in \Eq{eq:ordered_TCF} is considered, since only the feature on the positive part of the frequency axis is important for absorption spectroscopy.

\subsection{Vibronic Spectra via imaginary time PI methods}
\label{sec:vibronic_spectra}

\begin{figure}
\includegraphics{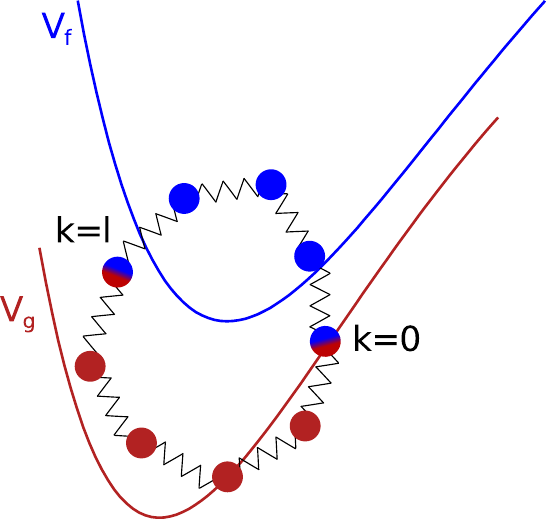}
\caption{\label{fig:sketch} Sketch of the effective ring polymer potential, \Eq{eq:RP_pot} for the case $P=9$ and $l=4$, where the 0-th and the $l$-th bead are marked.
The PESs $V_g$ and $V_f$ are shown in red and blue, respectively.
The color of the beads that ''feel`` one of the PESs is chosen accordingly.
}
\end{figure}

In the spirit of RPMD techniques, it is useful to consider the quantity of interest, $C_{\l}(t)$, at time zero for any given $\l$.~\cite{Craig-JCP-2004}
To reiterate, only the summand that lead to absorption spectra is kept in the expression for $C_{\l}(t)$, \Eq{eq:ordered_TCF}.
The nuclear trace therein is evaluated in the coordinate representation and the shift in imaginary time is equidistantly discretized, i.e.\ $\l=l \beta/P$ , where $l\in [0,P]$ and $P>0$ is a natural number which will become later the number of beads in the ring polymer.
The result has the form of a configuration-space average
\begin{align}
\label{eq:shifted_tcf_0}
C_{\l}(0)=\frac{1}{Z} \int \diff \mVec{R}_0 \mBra{\mVec{R}_0} \e^{-(P-l)\beta\mOp{H}_g/P}\mOp{D}^{g}_f \e^{-l\beta\mOp{H}_f/P} \mOp{D}^{f}_g \mKet{\mVec{R}_0}
\enspace ,
\end{align}
where $\mVec{R}$ describes the positions of all the nuclei in the system;
note that since $l$ is a discretized version of $\l$ we use both interchangeably to simplify the notation.
Following the standard imaginary time path integral approach,~\cite{Chandler-JCP-1981,Tuckerman-Book} each exponential term $\exp(-j\beta\mOp{H}_a/P)$, where $a=g,f$ and $j=P-l,l$ can be written as a product of $j$ identical factors $\exp(-\beta\mOp{H}_a/P)$.
Further, in total $P-2$ spatial closures $\int \diff \mVec{R}_k \mKet{\mVec{R}_k}  \mBra{\mVec{R}_k}=1$ with $k=1,\ldots l-1$ and $k=l+1,\ldots P-1$ for $a=f$ and $a=g$, respectively, are inserted in between those factors.
One additional closure with $k=l$ is put right next to $\mOp{D}^{g}_f$ yielding the corresponding eigenvalue $\mOp{D}^{g}_f\mKet{\mVec{R}_l} = {D}^{g}_f(\mVec{R}_l) \mKet{\mVec{R}_l} $ as it happens for $\mOp{D}^{f}_g\mKet{\mVec{R}_0}$ as well.
The resulting matrix elements $\mBra{\mVec{R}_{k+1}} \exp(-\beta\mOp{H}_a/P)  \mKet{\mVec{R}_{k}}$ are approximated via the symmetric Trotter factorization in order to separate position- and momentum-dependent terms.
The former result in the respective eigenvalues, whereas the latter are supplied by the momenta closures leading after some straightforward algebra to the well-known kinetic (spring) terms.
These terms stand for harmonic springs connecting adjacent beads of the resulting ring polymer; note that they are state-independent and thus coincide for $\mOp{H}_g$ and $\mOp{H}_f$.
%
Finally, the value $C_{\l}(0)$ gets the form of a configuration integral over the ring polymer coordinate space
\begin{align}
\label{eq:config_int}
C_{\l}(0)\approx
\frac{1}{Z} \int \diff \mMat{R}\, \e^{-\beta U_{l}(\mMat{R})} {D}^{g}_f(\mVec{R}_l) {D}^{f}_g (\mVec{R}_0)  \enspace,
\end{align}
where $\mMat{R}=\T{(\mVec{R}_0,\ldots\mVec{R}_{P-1})}$ is the ring polymer configuration obeying the cyclic condition $\mVec{R}_{P}=\mVec{R}_{0}$ and the effective ring polymer potential
\begin{align}
\label{eq:RP_pot}
 U_{l}(\mMat{R}) = K(\mMat{R})+ \frac{1}{P} \left [\sum_{k=0}^{l} \eta_k V_f(\mVec{R}_k)+ \sum_{k=l}^{P} \eta_k V_g(\mVec{R}_k) \right ] \enspace,\\
K(\mMat{R}) =\sum_{k=0}^{P-1}\frac{P}{2\beta^2\hbar^2} \T{(\mVec{R}_k-\mVec{R}_{k+1})} \mMat{M} (\mVec{R}_k-\mVec{R}_{k+1}) 
\enspace .
\end{align}
Here $K$ denotes the kinetic spring term, $\mMat{M}$ is the nuclear mass matrix and $\eta_k$ is equal to $1/2$ if $k$ corresponds to the first or the last summand, to $0$ if there is only one summand, which is the case if $l=0,P$, and to $1$ in all other cases.
Importantly each value of $l$ defines a particular PES and thus a particular \textit{realization} of the ring polymer, undergoing different dynamics as will become clear later, see \Eq{eq:final}.
An example of such a realization of the ring polymer is illustrated in \Fig{fig:sketch}.
One sees that there are two sets of beads, which ``feel'' either the upper or the lower PES as is illustrated by the blue or the red color, respectively.
Note that there are two ``boundary'' beads that distinguish one set of beads from the other and are influenced by the averaged potential (hence depicted with both colors).
The presence of the two distinguishable sets of beads breaks the cyclic symmetry of the ring polymer.
Remarkably, in the classical limit, $P=1$, $l$ can be equal to 0 and 1 and thus there would be \textit{two} realizations of the ring polymer according to \Eq{eq:RP_pot}.
This can be viewed as a consequence of the broken cyclic symmetry of the ring polymer; note that when the PESs are the same, the two potential energy terms in \Eq{eq:RP_pot} coincide yielding the standard PIMD expression for the ground state dynamics.
Equation~\ref{eq:config_int} remains exact in the limit $P\to \infty$.

In order to evaluate the configuration integral in \Eq{eq:config_int} correctly via the standard sampling methods, the partition function that appears in the expression has to normalize the density $\exp[{-\beta U_{l}(\mMat{R})}] $.
Thus we have to multiply and divide the expression by 
\begin{align}
Z_\l=\Tr \left [\e^{-(\beta-\l)\mOp{H}_g}\e^{-\l \mOp{H}_f} \right ] \propto
\int \diff \mMat{R}\, \e^{-\beta U_{l}(\mMat{R})}
\end{align}
leading to a prefactor $\xi_\l\equiv{Z_\l}/{Z}$ that has to be calculated.
As it is derived in the Supplement, this prefactor can be conveniently and still numerically exactly extracted as
\begin{align}
\label{eq:prefactor}
\xi_\l=\frac{\exp \left [ -\intop_0^{\l} \mAve{\mOp{V}_f-\mOp{V}_g}_\mu \diff \mu \right ]}{ 1 + \exp  \left [ -\intop_0^{\beta} \mAve{\mOp{V}_f-\mOp{V}_g}_\mu \diff \mu \right ]}
\enspace ,
\end{align}
where the average is defined as
\begin{align}
\mAve{\bullet}_\l=  \frac{1}{Z_\l} \Tr \left [\e^{-(\beta-\l)\mOp{H}_g}\e^{-\l \mOp{H}_f} \bullet\right ]
\enspace . 
\end{align}

Finally, putting together all the obtained results leads to a relation for $\bar{C}_w(0)$
\begin{align}
\label{eq:barCw}
\bar{C}_w(0) \approx
\frac{1}{P} \sum_{l=0}^{P} \eta_l w(\l)  \xi_{\l} \mAve {{D}^{g}_f(\mVec{R}_l) {D}^{f}_g (\mVec{R}_0)}_{\l}
\enspace,
\end{align}
where the imaginary time integration in \Eq{eq:Cwt} has been discretized via the trapezoidal rule.
It becomes apparent at this point that one has to independently simulate each summand in \Eq{eq:barCw}, which corresponds to the respective realization of the ring polymer defined by a particular value $\l$.
The factor $w(\l)$ provides an external weight to the realizations and can be chosen arbitrarily, whereas
the factors $\xi_\l$ are \textit{intrinsic} weights that are dictated by quantum statistical mechanics;
note that we consider distinguishable particles only.

The remaining question is how to approximate the nuclear dynamics, in particular, how to estimate the non-classical time evolution in \Eq{eq:shifted_tcf_BO}, i.e.\ $\exp[-\i\mOp{H}_f t / \hbar] \mOp{D}^{f}_g  \exp[\i\mOp{H}_g t / \hbar]$.
First, an effective Hamiltonian $\mOp{H}_\l$ is defined for each point in the imaginary time $\l=l\beta/P$ that corresponds to the effective potential $U_l(\mMat{R})$, \Eq{eq:RP_pot}.
Second, the Hamiltonian of the $a$-th state is rewritten as $\mOp{H}_a=\mOp{H}_\l+\mOp{H}_a-\mOp{H}_\l$ for $a=g,f$ in order to switch to the interaction representation~\cite{Kuehn-Book,Mukamel-Book} that yields
\begin{align}
\e^{-\im \mOp{H}_a t /\hbar}=\e^{-\im \mOp{H}_\l t / \hbar}\exp_{+} \left \{-\frac{\i}{\hbar}\int_0^{t}[ \mOp{H}_a(\tau)-\mOp{H}_\l ] \diff \tau \right \} \end{align}
where the time arguments represent a time evolution with respect to $\mOp{H}_\l$.
Third, the dynamics induced by $\mOp{H}_\l$ are approximated by the quasi-classical dynamics of the ring polymer with respect to
\begin{align}
H_\l(\mMat{R},\mMat{P})=\frac{1}{2P} \T{\mMat{P}} \mMat{M}^{-1} \mMat{P} + U_l(\mMat{R})
\end{align}
and the operators are replaced by their classical counterparts.
The momenta of the ring polymer $\mMat{P}$ are introduced as conjugate variables to the coordinates $\mMat{R}$ in the usual fashion.
Finally, the desired TCF is approximated as
\begin{align}
\label{eq:final}
\nonumber
&\bar{C}_w(t)\approx
\frac{1}{P} \sum_{l=0}^{P} \eta_l w(\l) \xi_\l\\
& \mAve {{D}^{g}_f(\mVec{R}_l) {D}^{f}_g (\mVec{R}_0(t)) \e^{\im / \hbar \int_0^{t}[ V_f(\mVec{R}_0(\tau))-V_g(\mVec{R}_0(\tau))] \diff \tau }}_\l
\enspace .
\end{align}
Equation~(\ref{eq:final}) is the main theoretical result of this work.
%
It should be stressed that this approximation leaves the density stationary at all times and excludes problems such as the infamous zero-point energy leakage.~\cite{Habershon-JCP-2009}
Additionally, this stationarity enables averaging along trajectories, on top of the averaging with respect to the initial conditions thereby greatly improving the statistical convergence.
The generalization to a larger number of states amounts to considering each transition separately according to \Eq{eq:final} and summing the results over.

\begin{figure*}
\centering
\begin{widetext}
\begin{minipage}{\textwidth}
\includegraphics[scale=0.9]{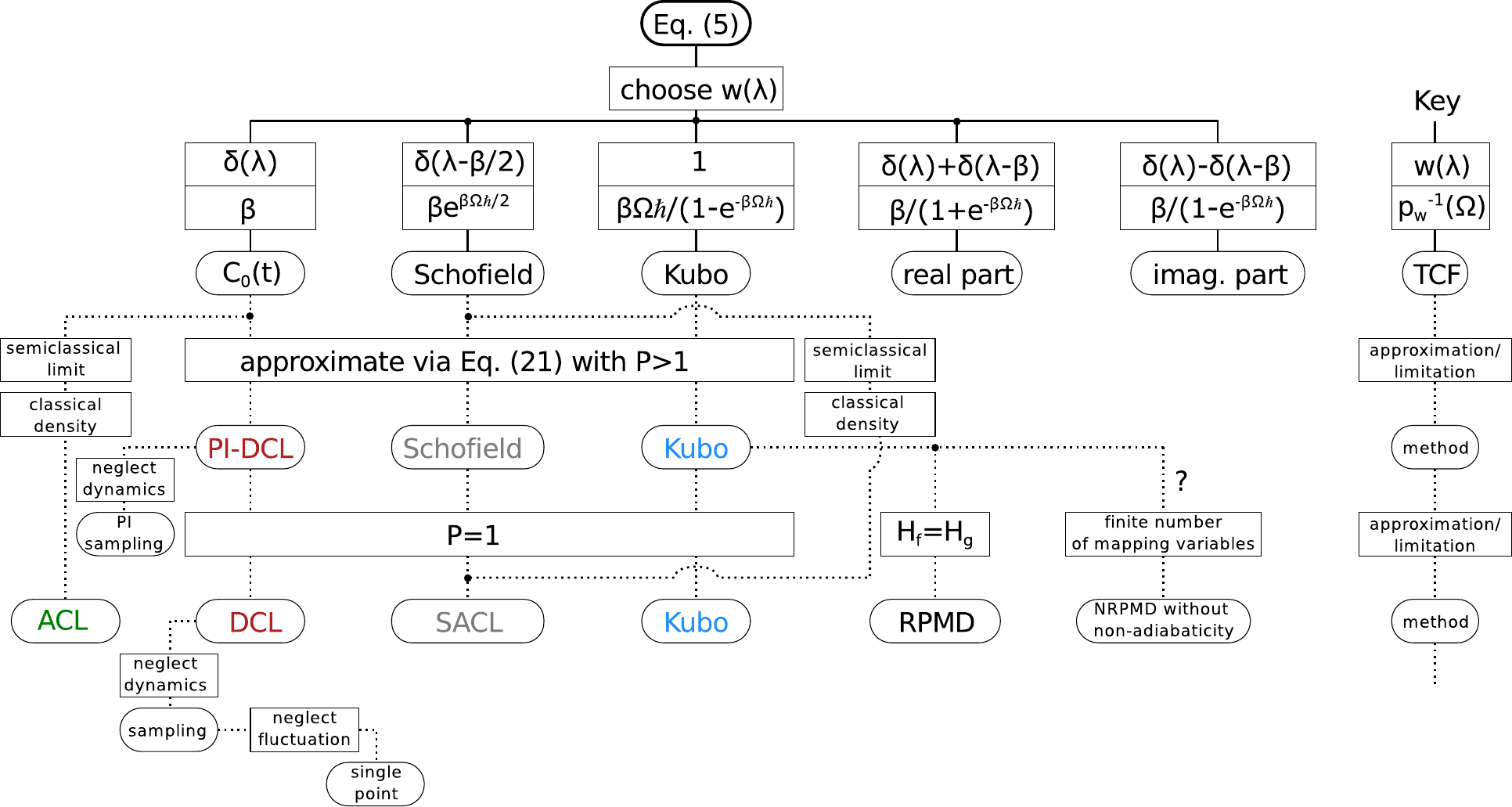}
\caption{\label{fig:overview}
Sketch of the developed formalism with several examples of the weighting function $w(\l)$ that lead to the well-established TCFs.
The legend on the right hand side deciphers the employed convention.
}
\end{minipage}
\end{widetext}
\end{figure*}

\subsection{Limiting cases}
\label{sec:limiting_cases}
%
The formalism presented in the previous section as well as several important limiting cases are sketched in \Fig{fig:overview}.
The starting point is \Eq{eq:main}, where a specific choice of the, in principle, arbitrary weighting function, $w(\l)$,  is made.
This choice defines the \QCF\ and leads to a particular TCF, including most of the well-established ones.
For instance, choosing $w(\l)=\delta(\l)$, see left column therein, defines a constant \QCF\ equal to $\beta$ and corresponds to the standard dipole autocorrelation function, \Eq{eq:dipole_TCF}.
Approximating the dynamics according to \Eq{eq:final} would lead to the dynamics with respect to the ground state \textit{only}, as it is done in the DCL, but taking nuclear quantum effects into account, hence termed PI-DCL.
The pure DCL limit can be straightforwardly obtained by setting the number of beads $P=1$.
The neglect of dynamics leads to sampling approaches, and further sacrifice of nuclear DOFs as such results in single point calculation methods as it was discussed in the Introduction.

For the case $w(\l)=\delta(\l)$ there exists also a completely different route to approximate the dynamics,~\cite{Shemetulskis-JCP-1992} starting from \Eq{eq:shifted_tcf_BO} with $\l=0$ in the Wigner representation and taking the semiclassical limit.
This results in the ACL dynamics, that is, the dynamics on the averaged PES $U_{1/2} \equiv 1/2  [ V_g + V_f]$.
In contrast to the approximation suggested in this work, the initial conditions for the ACL are sampled with respect to the PES of the ground state, $U_{0}$, thereby making the density non-stationary.
The presence of these non-equilibrium dynamics causes unphysical negativities in the spectrum as it is discussed in \Sec{sec:weighting_functions} and is proven in the Supplement.

Interestingly, the same approximation to the dynamics can be obtained within the presented formalism, by setting $w(\l)=\delta(\l-\beta/2)$, see second column in \Fig{fig:overview}, which results in the Schofield TCF,~\cite{Schofield-PRL-1960} being another TCF that has the desired symmetry properties for quasi- or semiclassical approximations.~\cite{Monteferrante2013,Bonella-JCP-2014}
Then the $P \to 1$ limit of the corresponding ring polymer potential can be interpreted as the averaged PES, $U_{1/2}$, defined above.
Thus, the dynamics in this case would be identical to the ACL one with the big advantage that the density is stationary.
Since it stems from the Schofield TCF, it is referred to as Schofield ACL (SACL) in the following.
It is worth mentioning that analogous to ACL, one can obtain SACL by starting from the Schofield TCF and following the semiclassical route described in Ref.~\onlinecite{Shemetulskis-JCP-1992}.

Choosing $w(\l)=1$, which corresponds to a democratic average over all possible ring polymer realizations, results in the Kubo TCF.
We believe that \Eq{eq:final} for this choice of the weighting function constitutes the limit of infinite number of mapping variables for the NRPMD method~\cite{Richardson-JCP-2013,Richardson-CP-2016} in the \textit{adiabatic} regime; note that a similar suggestion was made in Ref.~\onlinecite{Richardson-CP-2016}.
Further, setting $\mOp{H}_f=\mOp{H}_g$ yields the standard Kubo-transformed TCF, which serves as the basis for the state-of-the-art RPMD method.~\cite{Craig-JCP-2004,Ceriotti-CR-2016}
Note that the real as well as the imaginary part of $C_0(t)$ can be obtained by setting $w(\l)=\delta(\l)\pm\delta(\l-\beta)$, respectively.

\section{Computational Details}
\label{sec:Computational_Details}
In order to probe a chemically relevant regime, the presented protocol was applied to a diatomic which mimics the OH bond of a gas-phase water molecule.
The PESs for the states, $V_g$ and $V_f$, were represented by a Morse potential $V_a(r)=E_a (1-\exp[{-\alpha_a(r-\deq_a)}] )^2$, where the ground state parameters  $E_g=0.185$\,au, $\alpha_g=1.21$\,au$^{-1}$ and $\deq_g=1.89$\,au  were taken from the qSPC/Fw water model and $r$ is the distance between O and H atoms.~\cite{Paesani-JCP-2006}
To have non-trivial spectra, the frequency and the position of the minimum for the (final) excited state PES were chosen to be different from the ground state ones, whereas the dissociation energy, was set the same $E_f=E_g\equiv E$.
The particular values for the parameters $\alpha_f$ and $\deq_f$ for the cases considered are given in the results section and the respective PESs are sketched in the insets of \Fig{fig:sys}.
Further, the Condon approximation for the dipoles was used and the spectra were broadened with a Gaussian function with the dispersion $\sigma=0.002$\,au to account for dephasing due to the interactions with an environment.

Spectra were simulated according to \Eq{eq:final}, with 500 uncorrelated initial conditions for each summand therein (that is each realization of the ring polymer with respect to $\l$) employing MD with a Langevin thermostat at two different temperatures; ACL required 5000 trajectories due to non-stationarity of the dynamics.
One was the ambient $T=300.0$\,K corresponding to $\beta\hbar\omega_g=18.6$, where $\omega_g$ is the harmonic frequency of the ground state.
The other higher $T=1117.6$\,K implied $\beta\hbar\omega_g=5$.
As a side product of this sampling, the prefactors $\xi_\l$ were obtained via the gap average $\langle \mOp{V}_f-\mOp{V}_g \rangle_\l$ according to \Eq{eq:prefactor}.
An MD trajectory according to $H_\l$ with a length of $12000.0$\,au and a timestep of $3.0$\,au was performed using the Velocity-Verlet algorithm, starting from each initial condition.
The time evolution resulting from the kinetic spring term has been performed analytically as described in Ref.~\onlinecite{Ceriotti_JCP_2010}.
The desired TCF was obtained as an ensemble average, that is as an average over a swarm of initial conditions and, owing to the stationarity of the density, over time, that is along the trajectories.
Exact results were obtained in the basis of 50 eigenstates of the harmonic oscillator with the frequency $\omega_g$.

\section{Results and Discussion}
\label{sec:Results}

\subsection{Common weighting functions}
\label{sec:weighting_functions}

\begin{figure*}
\centering
\begin{widetext}
\begin{minipage}{\textwidth}
\includegraphics[scale=0.9]{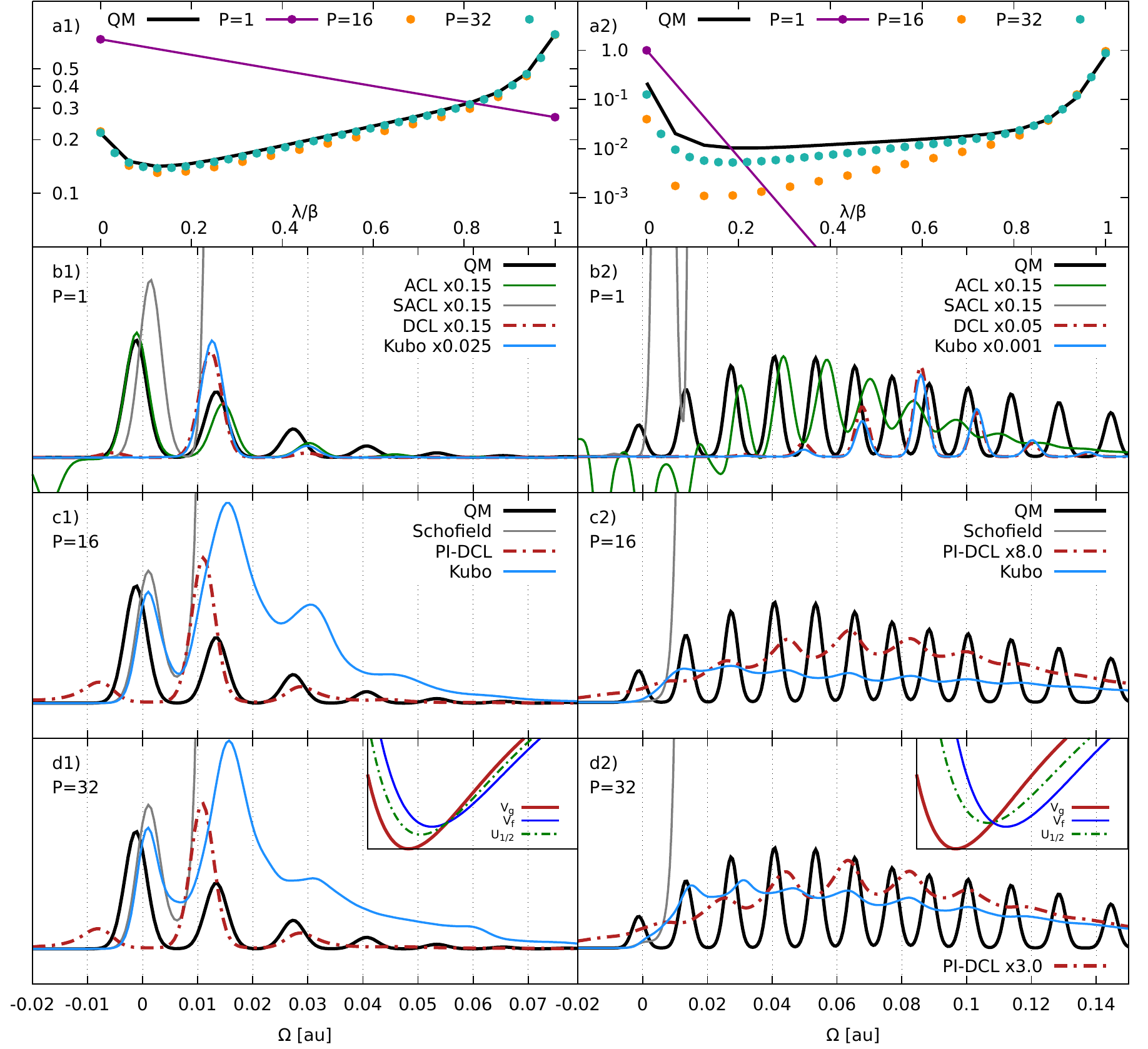}
\caption{\label{fig:sys}
The two-level system with the parameters $\alpha_f/\alpha_g=0.86$ at 300\,K, left: $\deq_f-\deq_g=0.22$\,au and right: $\deq_f-\deq_g=0.5$\,au, see text.
Panels a) Intrinsic weights, $\xi_\l$, b)-d) absorption spectra for $P=1,16,32$, respectively.
Potential energy surfaces are sketched in the insets.
}
\end{minipage}
\end{widetext}
\end{figure*}

The performance of all the aforementioned methods, namely PI-DCL, DCL, ACL, Schofield, SACL and Kubo, is compared against the numerically exact reference in \Fig{fig:sys} for the two shifted Morse oscillators, see insets therein.
The shifts were chosen to be $\deq_f-\deq_g=0.22$\,au(left) and $\deq_f-\deq_g=0.5$\,au (right), with a moderate ratio $\alpha_f/\alpha_g=0.86$ and using ambient temperature.
In panel a1) therein the convergence of intrinsic weights, $\xi_\l$, to the exact result is illustrated.
To start, the weights for $P=1$ (classical case) are qualitatively wrong.
In this case there are only two points, corresponding to the two realizations of the classical particle being \textit{either} in the ground \textit{or} in the excited state, see \Sec{sec:Theory} for a discussion.
Importantly, despite being an exponential quantity, the intrinsic weights can be obtained with sufficient accuracy with $P=16$ and completely converge at $P=32$; note the log scale.
These numbers of beads are typical for reaching convergence for ground-state properties of water at ambient conditions.~\cite{Habershon-JCP-2009,Ivanov-JCP-2010}
It is natural to expect the same convergence behavior, when the shift between PESs is relatively small.

Switching to the absorption spectra depicted in \Fig{fig:sys}b1)-d1), one sees that the exact solution reveals a typical Franck-Condon progression with a Huang-Rhys factor below $0.5$, which implies the maximal intensity at the 0-0 transition.~\cite{Kuehn-Book,Schroeter-PR-2015}
For $P=1$, panel b1), DCL and Kubo results are very similar, since $\xi_0$ is much higher than $\xi_\beta$, see panel a1), and thus the contributions stemming from the ground state play the most important role.
Both methods fail completely to reproduce the exact spectrum in this parameter regime.
In particular, the intensities are dramatically overestimated (note the scaling factor) and
the maximum is not at the exact 0-0 transition.
The vibronic progression is significantly suppressed and features wrong frequencies and almost symmetrical shapes, as it was discussed for DCL before.~\cite{Karsten-JPCL-2017,Karsten-JCP-2017}
The shape is more symmetrical for DCL than for Kubo, because the Kubo \QCF, being responsible for the detailed balance, suppresses the signal below zero frequency; the \QCFs\ are listed in \Fig{fig:overview} and the non-trivial ones are plotted in \Fig{fig:improved}b).
In contrast, ACL results satisfactorily agree with the exact ones, apart from a slight difference in the fundamental frequency.
Also the infamous negativities, which are intrinsic to the method can be observed to the left of the 0-0 transition.
Finally, the spectral intensities resulting from the Schofield function (SACL) grow uncontrollably due to the \QCF.
To illustrate, the \QCF\ that reads $\exp[\hbar\beta\w/2]$, would yield approximately 3000, $9\cdot10^6$ and $2.7\cdot10^{10}$ for the first, second and third overtones of the characteristic frequency $\omega_f$, respectively; note that whereas the frequency is growing linearly with the vibrational quantum number in vibronic progression, the \QCF\ grows exponentially.
This suggests that a more suitable choice of the \QCF\ can yield better numerical results, as is discussed in \Sec{sec:improvements}.

When it comes to more beads, that is $P=16$ in panel c1), PI-DCL exhibits reasonable amplitudes but still yields fairly symmetric wrong spectral shapes.
This is a manifestation of the fact that the ground-state dynamics only cannot reproduce the spectra which are significantly dependent on the peculiarities of the excited state.
The Kubo results improve a lot with respect to the spectral structure, as they include contributions from the beads evolving on the PES of the excited state, included due to the intrinsic weights in the proper way.
The shape is still not correct due to over-pronounced contributions from large values of $\l$, which in principle can be healed by a proper choice of the \QCF, see \Sec{sec:improvements}.
Schofield spectra still grow uncontrollably with frequency due to the \QCF, which does not depend on the number of beads.

All these findings remain the same when the spectra are converged with respect to the number of beads, panel d1), as it could be already anticipated from the small difference of the intrinsic weights for $P=16$ and $P=32$ shown in panel a1).

In order to test a different regime, where the vibronic progressions are more pronounced, the same system with the displacement of the PESs increased to $\deq_f-\deq_g=0.5$\,au was considered, see right panels in \Fig{fig:sys}.
Following the same structure of the discussion, the curve for intrinsic weights is
an order of magnitude lower, is much steeper at the imaginary time borders and converges slower to the exact result.
The latter is intuitively expected, since a larger number of beads is needed to account for the increased displacement of the PESs as can be understood from~\Fig{fig:sketch}.
The difference between $\xi_0$ and $\xi_\beta$ for $P=1$ suggests that the results for Kubo and DCL would match even closer than it was observed in panel b1).
Indeed, the difference between the two is barely visible, see panel b2); note that they are still not identical due to the different \QCFs.
Again, both show dramatically overestimated intensities and rather symmetric lineshapes, whereas
the ACL spectra are qualitatively better, though suffer increasingly from the negativities.
When the number of beads is increased to $P=16$, panel c2), the PI-DCL amplitude becomes very low due to the underestimated intrinsic weight at $\l=0$.
The Kubo spectrum improves over that from DCL but the peaks are still much broader than in the QM reference and are not at the correct positions.
These trends are preserved when the beads' number is increased further ($P=32$, panel d2)).
The PI-DCL amplitudes are getting more reasonable supporting the notion of the converged intrinsic factors, but the shape and peak positions still do not fit to the QM result.
The quality of Kubo spectra improves further and the Schofield spectra diverge again.
Practically, there is no qualitative difference in terms of the performance of each method for the two regimes considered.

\begin{figure*}
\includegraphics[scale=0.9]{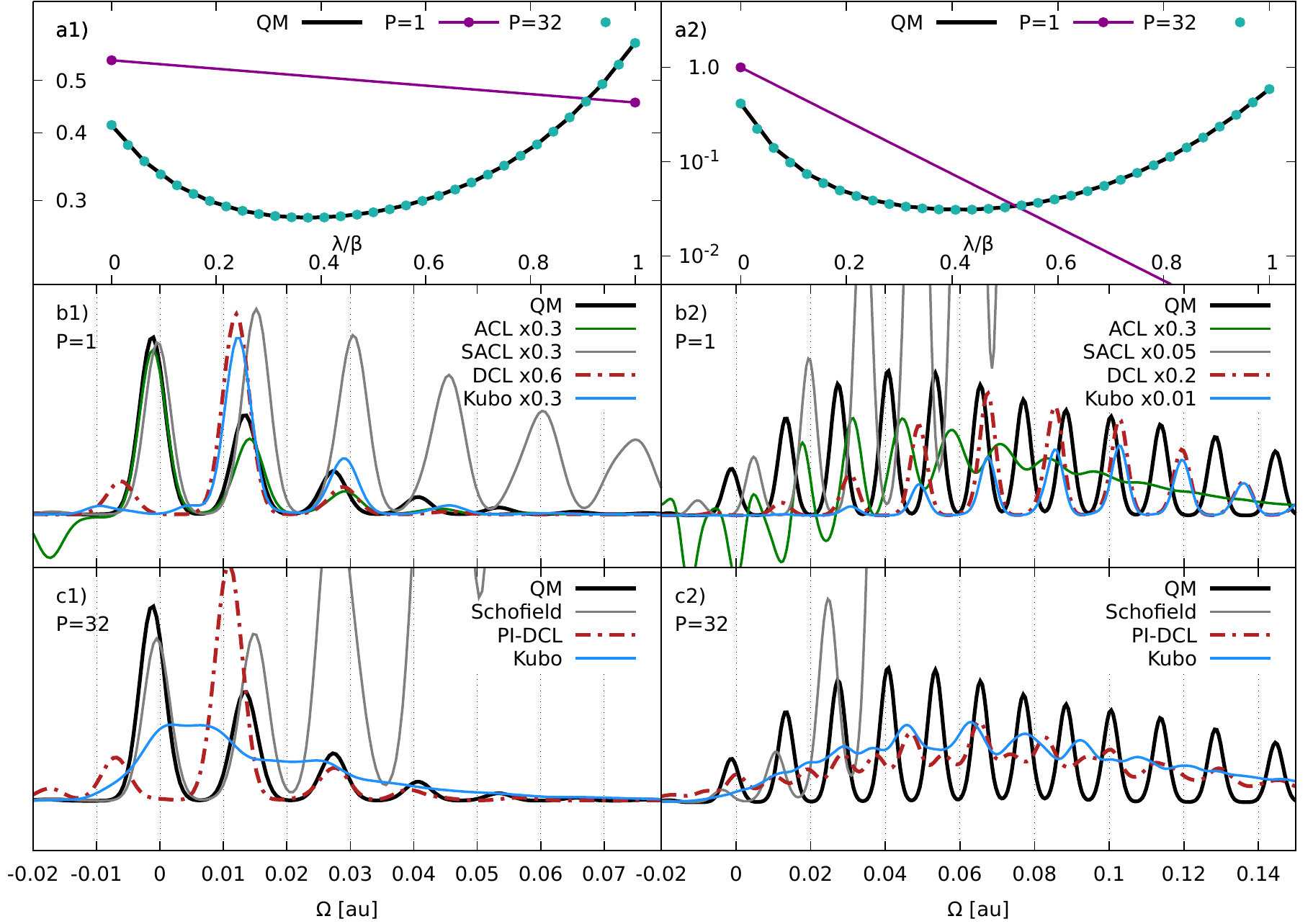}
\caption{\label{fig:sys2}
The structure of the figure coincides with that of \Fig{fig:sys} with the only difference being the higher temperature $T=1117.6$ ($\beta\hbar\omega_g = 5$)
}
\end{figure*}

Yet another possible regime can be accessed via increasing the temperature (or, equivalently, by decreasing the frequencies) while keeping the (other) parameters of the ground and excited PESs the same, see \Fig{fig:sys2}.
Starting with the system with moderate shift in the left panels one can see that the intrinsic weights, $\xi_\l$, are converged.
Switching to the spectra in panel b1) one sees that DCL and Kubo again fail to describe the exact spectrum.
The ACL spectrum reveals small negativities and slightly incorrect peak positions but overall demonstrates the best agreement among the methods considered.
Interestingly in this regime the Schofield spectrum remains bound which can be explained by a more moderate \QCF\ as a consequence of the higher temperature.
However the intensities are still strongly overestimated especially for higher frequencies.

While the number of beads is increased, panels c), the PI-DCL results do not improve over pure DCL ones although the amplitudes become better.
The Kubo spectrum loses the fine structure whereas the envelop resembles the shape of the exact result.
The Schofield spectrum again becomes unbound as it was for the lower temperature.

The conclusions for the other system are the same apart from the exploding SACL spectra in panel b2) as the interested reader can figure out. 

\subsection{Possible improvements}
\label{sec:improvements}
\begin{figure*}
\includegraphics[scale=0.9]{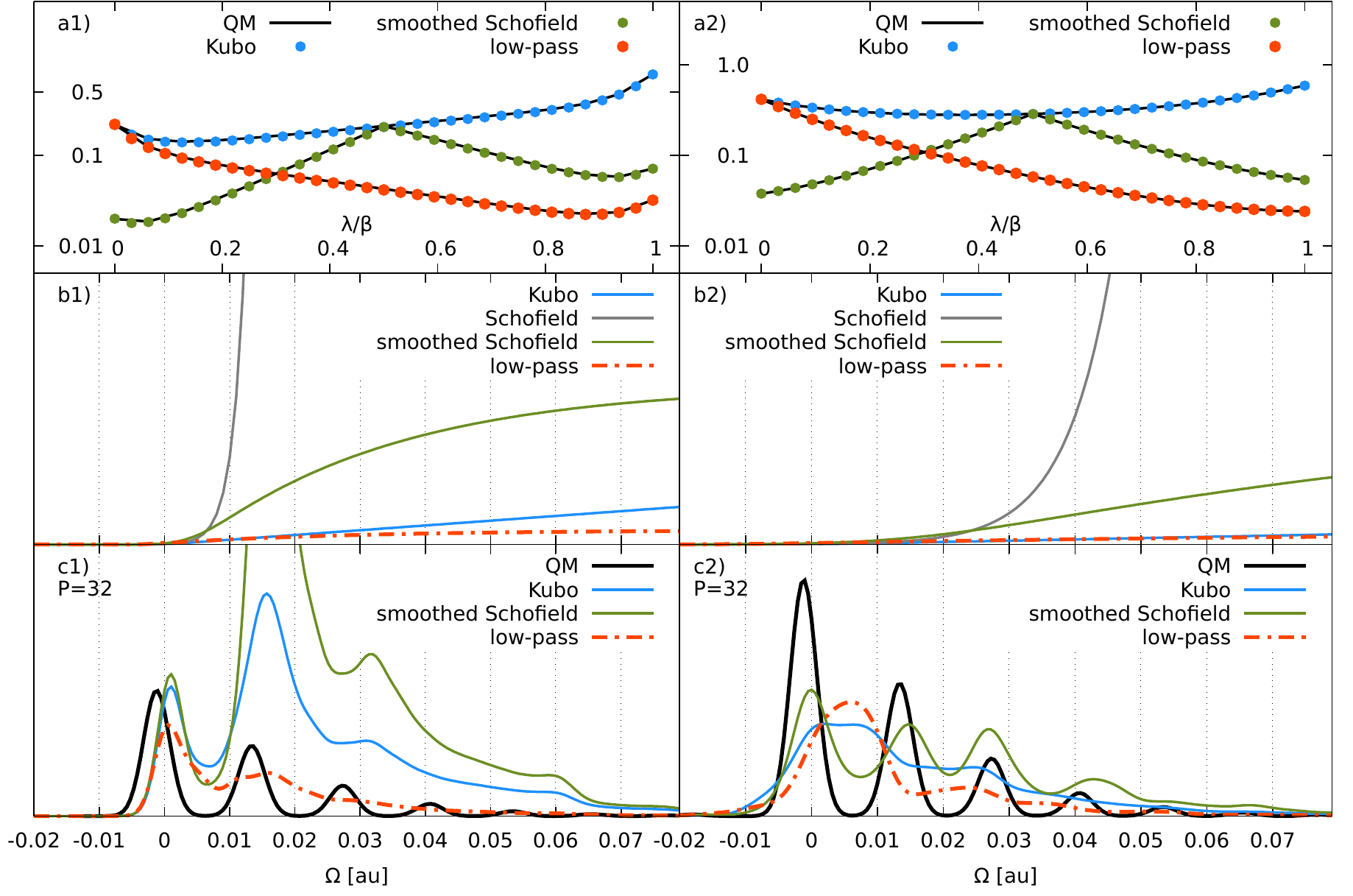}
\caption{\label{fig:improved}
The two-level system with the parameters, left: $\alpha_f/\alpha_g=0.77$, $d_f-d_g=0.17$\,au at $T=300$\,K and 
right:  $\alpha_f/\alpha_g=0.86$, $d_f-d_g=0.22$\,au at $T=1117.6$\,K, see text. 
The parameters for the latter coincide with that of left panels in \Fig{fig:sys}.
The smoothing for the Schofield function $\beta\eps=4.8$, and for the low-pass filter $\beta\eps=3.2$.
Panels a) Filtered intrinsic weights, $\xi_\l w(\l)$, b) \QCFs, c) absorption spectra for $P=32$.
}
\end{figure*}

As it was shown in the previous section, ACL performs generally quite well with respect to the lineshapes, but reveals negativities due to non-equilibrium dynamics, which is a rather severe problem.
The equilibrium version of ACL, which does not have this deficiency, stems from Schofield correlation function, hence termed SACL in \Sec{sec:limiting_cases}.
Unfortunately, the respective \QCF\ is enormously large in the relevant frequency regions (\Fig{fig:improved}b)) thereby leading to a numerical instability (large times small number) for spectra.
It is thus desirable to preserve the dynamics of SACL having a more ``gentle'' \QCF.
To reiterate, in contrast to intrinsic weights $\xi_\l$ dictated by statistics, external weighting function $w(\l)$ can be chosen arbitrarily.
One possibility is to choose
\begin{equation}
w(\l)=\e^{-|\l-\beta/2| \epsilon}
\enspace,
\end{equation}
which ``smooths'' the delta peak around $\beta/2$ (corresponding to the Schofield TCF), with the smoothing being controlled by the value of $\epsilon$.
It can be clearly seen in \Fig{fig:improved}b) that the resulting \QCFs\ (green curves) indeed exhibit a significantly more moderate growth with frequency.
Another possibility logically follows from the conclusion that Kubo performs reasonably well, but suffers from over-pronounced contributions with large $\l$, thus, suggesting
\begin{align}
w(\l)=\e^{-\l \epsilon}
\enspace .
\end{align}
This constitutes a low-pass filter that suppresses the unwanted contributions to spectra.
The respective filtered intrinsic weights, $\xi_\l w(\l)$, are shown in the upper panels of \Fig{fig:improved}.
One sees that the smoothed Schofield includes all the contributions keeping the emphasis on the middle one,
whereas the low-pass filter indeed suppresses the contributions with large $\l$.
To reiterate, Kubo and Schofield can be viewed as two limiting cases, the former taking all the contributions into account democratically and the latter picking only a single ($\l=\beta/2$) contribution.

The respective spectra for the system with the small shift are shown in the lower panels of \Fig{fig:improved};
the results for the other system are given in the Supplement.
One sees that at lower temperatures (larger $\beta$), panel c1) the low-pass filter appears to be the better choice.
In particular, it removes the over pronounced high frequency contributions the Kubo results suffer from, whereas the smoothed Schofield filter emphasizes these contributions even more.
However, the low-pass filter results in the dynamical features that are washed out with respect to the exact spectrum.
In contrast, for the higher temperature case low-pass performs as badly as the Kubo does, whereas the smoothed Schofield filter reveals the fine spectral structure with a decent quality.
In both cases the smoothed Schofield spectra stay bound, as it is implied by the choice of the external weight.
Unfortunately the suggested weights did not improve the results for the system with the larger shift, see Supplement.

To summarize, the present study suggests that a non-standard form of $w(\l)$ \textit{can} be beneficial in comparison to common choices with respect to quality and numerical stability.
However, a universal recipe for choosing the weighting function is hard to formulate for the general case and requires further investigations.

\section{Conclusions and outlook}\label{sec:conclusions}

The central question of this work was to determine the optimal starting point for approximate quasi-classical protocols aiming at vibronic spectroscopy.
To reiterate, the Kubo-transformed TCF enjoyed success in describing the ground state dynamics, due to fundamental symmetry properties that make it the most classical-like quantum TCF.~\cite{Craig-JCP-2004, Ramirez2004}
Unfortunately, direct practical application of the Kubo TCF to \textit{vibronic} transitions leads to either numerical or conceptual deficiencies, namely it is either intractable numerically or becomes a complex function.
Therefore in the present work, we suggested a generalized form of quantum time correlation functions and several approximations based on it that can be used, e.g., for the linear vibronic absorption spectroscopy.
The generalization is done via employing a shift in imaginary time and introducing a weighting function,
thereby uniting many TCFs that have been reported in the literature before as well as many other yet not considered variants.

Absorption spectra resulting from various weighting functions and thereby from different TCFs were compared against the exact results, which can be straightforwardly obtained for 1D two-level model system considered here.
%
As it was already shown in the previous works,~\cite{Karsten-JPCL-2017, Karsten-JCP-2017} DCL-based methods typically exhibit (incorrectly) symmetrical spectral shapes with possibly wrong frequencies, when the ground and excited PESs differ significantly.
Apart from the negativities which stem from the non-equilibrium dynamics, the results of the ACL method agree well with exact spectra in terms of frequencies and intensity ratios in vibronic progressions.
We would like to stress that these negativities are not trivial artifacts that can be manually removed from the spectra.
A possible solution based on the Schofield TCF, that leads to an equilibrium dynamics, suffers from an unhealthy growth of the shift correction factor for both low- and high-temperature regimes.
The performance of the popular Kubo TCF is in most cases not superior with respect to the aforementioned ones that treat excited-state dynamics explicitly, exhibiting washed-out fine spectral structures and/or incorrect frequencies and intensities.
The results could be improved upon choosing a more complicated weighting function form, however staying unsatisfactory for large shifts between ground and excited state PESs.

To summarize, a successful method should employ dynamics, which accounts for the involved PESs simultaneously.
The imaginary-time PI approach not only incorporates nuclear quantum effects, but also yields a possibility to formulate a general form of the quantum TCF.
Approximations to the latter may provide more practical numerical protocols than those based on the state-of-the-art Kubo TCF.
Nonetheless, the presented formalism sacrifices the non-adiabatic effects; to allow for them NRPMD~\cite{Richardson-JCP-2013,Ananth-JCP-2013} is hitherto the method of choice.
We believe that the suggested methodology constitutes the limit of infinite number of mapping variables for the NRPMD method in the \textit{adiabatic} regime.

The existing freedom for the choice of the \QCF\ provides attractive possibilities for finding an optimal one suited for a particular problem as well as, if possible, for a class of problems or even for the general case.
Practically, this can be done, e.g., using machine learning techniques.
Importantly, such an optimization can be performed \textit{a posteriori} based on the existing dynamical results without recomputing them.
Ideally, this should be supported by a physical motivation that can specify the particular form of the weighting function, or explain the numerically obtained one.
This is clearly a project on its own, which requires a significant additional effort, and is a subject for a future research.

\section{Supplementary Material}

The supplementary material contains the relation between $S_\l(\w)$ and the absorption line shape $S_0(\w)$, the derivation of the expression for the intrinsic weights, \Eq{eq:prefactor}, the proof that the negativities in the spectra are due to non-equilibrium dynamics and the results of applying the filters to the system with the large PES shift.

\begin{acknowledgments}
We acknowledge financial support by 
the Deutsche Forschungsgemeinschaft (KU~952/10-1 (S.K.), IV~171/2-1 (S.D.I.)).
\end{acknowledgments}

\bibliography{./united}
\end{document}